# Discrete Fourier Transform Improves the Prediction of the Electronic Properties of Molecules in Quantum Machine Learning

Alain B. Tchagang, *IEEE, Member,* and Julio J. Valdés, *IEEE, Senior Member,*

*Abstract*—high-throughput approximations of quantum mechanics calculations and combinatorial experiments have been traditionally used to reduce the search space of possible molecules, drugs and materials. However, the interplay of structural and chemical degrees of freedom introduces enormous complexity, which the current state-of-the-art tools are not yet designed to handle. The availability of large molecular databases generated by quantum mechanics (QM) computations using first principles open new venues for data science to accelerate the discovery of new compounds. In recent years, models that combine QM with machine learning (ML) known as QM/ML models have been successful at delivering the accuracy of QM at the speed of ML. The goals are to develop a framework that will accelerate the extraction of knowledge and to get insights from quantitative process-structure-property-performance relationships hidden in materials data via a better search of the chemical compound space, and to infer new materials with targeted properties. In this study, we show that by integrating well-known signal processing techniques such as discrete Fourier transform in the QM/ML pipeline, the outcomes can be significantly improved in some cases. We also show that the spectrogram of a molecule may represent an interesting molecular visualization tool.

## I. INTRODUCTION

Finding new materials with desired properties is strategic to the innovation and progress of many industry sectors. Recent advances have shown that data-to-knowledge approaches are beginning to show enormous promise within materials science. Intelligent exploration and exploitation of the vast materials property space has the potential to alleviate the cost, risks, and time involved in trial-by-error approach experiment cycles used by current techniques to identify useful materials [1]. Additionally, data-driven approaches can also yield valuable insights into the fundamental factors underlying materials behavior and lead to the discovery of new rules for material design and synthesis [1]. To significantly accelerate the pace of discovery using such data-to-knowledge approaches, advanced machine learning and statistical signal processing techniques have to be developed [2].

The availability of huge molecular databases [3] obtained from quantum mechanics (QM) computations using first principles open new opportunities for the data mining, signal processing (SP), and machine learning (ML) communities to either develop novel or use well established computational tools to analyze and obtain fast solutions at the accuracy of computational QM [2-9]. In recent years, ML has been successfully applied to tackle some of these problems. An example is the prediction of the electronic properties of molecules, bypassing at the same time the resolution of complex QM equations such as the Schrödinger's equations.

The success of these QM/ML models relies on better molecular representations that can be used as input to the ML methods. These representations should be (i) invariant to transformations that do not change the property of molecules, in particular translations, rotations, and nuclear permutations; (ii) unique; (iii) continuous and differentiable [6]. Coulomb matrix representation and its variants have shown to provide such descriptors [4, 5, 6, 8, 9]. It is invariant to translation and rotation but not to permutations or re-indexing of the atoms. Methods to tackle this issue have been proposed [4].

After algebraic manipulations as described in the Methods Section, the Coulomb matrix becomes a 1-dimensional (1D) order numerical sequence representation of a molecule. From the SP perspective, it can be treated as a 1D signal. SP has a long tradition of dealing with signals, and its techniques have been successfully applied over years to preprocess, transform, analyze and extract useful information from a wide variety of signals: static and dynamic, discrete and continuous, stationary and non-stationary, and from diverse disciplines such as engineering, biology, physiology, medicine, geology, images, astronomy, economics, and social sciences [10-11]. In this study, we showed by means of an example that by analyzing the 1D signal representation of molecules in the frequency domain using the discrete Fourier transform, the outcomes of ML models are significantly improved in some cases.

To validate, our claim, we tackle the prediction of the electronic properties of molecules using their 1D signal. The solution to this problem is very important in QM. It will not only accelerate the discovery of new materials, but also improve the accuracy of computational QM based on first principles. A Gaussian kernel ridge regression is used to model the relationship between the input (i.e. the 1D signal) and the output (i.e. electronic properties of molecules) as has been done in the literature [4, 5, 7]. The results obtained show significant improvement when the 1D Coulomb signals is preprocessed using DFT.

The rest of this paper is organized as follows. In Section II, the dataset used in this study is described. Section III provides a detailed description of the proposed method. Section IV presents the results and Section V the conclusions.

Research supported by the National Research Council of Canada.

A. B. Tchagang and J. J. Valdés are with Digital Technologies Research Centre, National Research Council, Ottawa, ON K1A 0R6 Canada. Phone: 613-993-5716; Corresponding author: alain.tchagang@ nrc-cnrc.gc.ca

## II. MATERIALS

The QM7 dataset used in this study is a subset of the GDB-13 dataset [3]. The version used here is the one published in [9] consisting of 7102 small organic molecules and their associated atomization energy.

## III. METHODS

Using the atomic coordinates of each molecule as described in the QM7 dataset, its Coulomb matrix is computed. Next, the 1D signal representation of each molecule is extracted from its Coulomb matrix. A kernel Ridge regression model is trained to predict the electronic properties of molecules from their 1D signal representation.

### A. 1D Coulomb Signal: Input to the Predictor

The Coulomb representation has recently been widely used as molecular descriptors in the QM/ML models. Given a molecule its Coulomb matrix $c = [c_{ij}]$ is defined using **Equation 1**.

$$c_{ij} = \begin{cases} 0.5 Z_i^{2.4} & \text{for } i = j \\ \dfrac{Z_i Z_j}{\| R_i - R_j \|} & \text{for } i \neq j \end{cases} \quad (1)$$

Where $Z_i$ is the atomic number of atom i, and $R_i$ is its position in atomic units [9]. The Coulomb matrix is symmetric and has as many rows and columns as there are atoms in the molecule. It is invariant to rotation, translation but not to permutation of its atoms. One remedy is to sort these matrices by order with respect to the norm-2 of their columns and simultaneously permuting rows and columns accordingly. After the ordering step and given the symmetry of these matrices, it is customary to only consider their lower triangular part [4, 9], and to unfold them row-wise in a 1D vector of length L. The output of this process is a 1D finite numerical sequence. In this study, we will refer to it as the 1D Coulomb signal $y(m,:) = y_m[l]$, with $l = 1$ to $L$ and m a given molecule. For a set of M molecules, their 1D Coulomb signals can be organized in an M×L matrix y.

$$y = \begin{bmatrix} y_{11} & y_{12} & \ldots & y_{1l} & \ldots & y_{1L} \\ y_{21} & y_{22} & \ldots & y_{2l} & \ldots & y_{2L} \\ \vdots & \vdots & \ldots & \vdots & \ldots & \vdots \\ y_{m1} & y_{m2} & \ldots & y_{ml} & \ldots & y_{mL} \\ \vdots & \ldots & \ldots & \vdots & \ldots & \vdots \\ y_{M1} & y_{M2} & \ldots & y_{Ml} & \ldots & y_{ML} \end{bmatrix} \quad (2)$$

The $m^{th}$ row of $y$ represents the 1D Coulomb signal of the $m^{th}$ molecule. Given that molecules have different number of atoms, the short ones are padded with zeros so that all the 1D Coulomb signals have the same length (equal to the one with the largest number of atoms).

### B. Discrete Fourier Transform

The 1D Coulomb signal $y_m[l]$ is a discrete sequence of length L and can be analyzed using SP techniques such as DFT [10-11]. The DFT of $y_m[l]$ is another sequence $Y_m[k]$ of the same length L (k = 0 to L-1), providing a measure of the frequency content at frequency k, which corresponds to an underlying period of L/k samples, where the maximum frequency corresponds to k = L/2, assuming that L is even.

$$Y_m(k) = \sum_{l=0}^{L-1} y_m(l) \exp(-j 2\pi k l / L) \quad (3)$$

$$Y = \begin{bmatrix} Y_{11} & Y_{12} & \ldots & Y_{1l} & \ldots & Y_{1L} \\ Y_{21} & Y_{22} & \ldots & Y_{2l} & \ldots & Y_{2L} \\ \vdots & \vdots & \ldots & \vdots & \ldots & \vdots \\ Y_{m1} & Y_{m2} & \ldots & Y_{ml} & \ldots & Y_{mL} \\ \vdots & \ldots & \ldots & \vdots & \ldots & \vdots \\ Y_{M1} & Y_{M2} & \ldots & Y_{Ml} & \ldots & Y_{ML} \end{bmatrix} \quad (4)$$

For the M molecules, their DFT can also be expressed in a matrix form Y, where each row of Y corresponds to the DFT of the 1D Coulomb signal of a molecule. Note that in this study, unless specified, DFT refers to Discrete Fourier Transform. Not to confuse with Density Functional Theory.

### C. Electronic Properties of Molecules

The electronic properties of molecules are organized in an M×|P| matrix P.

$$P = \begin{bmatrix} p_{11} & p_{12} & \ldots & p_{1x} & \ldots & p_{1|P|} \\ p_{21} & p_{22} & \ldots & p_{2x} & \ldots & p_{2|P|} \\ \vdots & \vdots & \ldots & \vdots & \ldots & \vdots \\ p_{m1} & p_{m2} & \ldots & p_{mx} & \ldots & p_{m|P|} \\ \vdots & \ldots & \ldots & \vdots & \ldots & \vdots \\ p_{M1} & p_{M2} & \ldots & p_{Mx} & \ldots & p_{M|P|} \end{bmatrix} \quad (5)$$

|P| is the number of electronic properties and M the number of molecules. The $m^{th}$ row of p represents the electronic properties of the $m^{th}$ molecule. The entry $p_{mx}$ is a real number that corresponds to the $x^{th}$ property of the $m^{th}$ molecule, and it is obtained from computational QM first principles [4-9].

### D. Kernel Ridge Regression

The predictor takes as input the 1D Coulomb signal or its DFT transform, and outputs the desired electronic property. Kernel Ridge regression (KRR) is used in this study to model such relationship. It has previously been used by several authors for the predictions of electronic properties of molecules [4, 9] and for the approximation of density functional theories [7]. The central idea of kernel-based ML is to derive nonlinear versions of linear ML algorithms in a systematic way. This is done by mapping the inputs into a higher-dimensional space and applying the linear algorithm there [9]. Here we used the Gaussian kernel, and the Euclidean distance between molecules defines the Gaussian kernel matrix:

$$K(y_m, y_{m'}) = \exp\left(-\frac{\|y_m - y_{m'}\|^2}{2\sigma^2}\right). \quad (6)$$

In the DFT domain, we can also write:

$$K_{DFT}(Y_m, Y_{m'}) = \exp\left(-\frac{\|Y_m - Y_{m'}\|^2}{2\sigma_{DFT}^2}\right). \quad (7)$$

Where $\sigma$ and $\sigma_{DFT}$ are the kernel bandwidth and $Y_m$ = DFT($y_m$). K and $K_{DFT}$ are the kernel matrices. The estimated property $P_m^e$ of the $m^{th}$ molecule is computed as the sum over weighted Gaussians in the original domain or in the DFT domain as follows.

$$p_m^e = \sum_{m'} \beta_{m'} K(y_m, y_{m'}), \quad (8)$$

$$p_m^e = \sum_{m'} \beta_{m'}^{DFT} K_{DFT}(Y_m, Y_{m'}). \quad (9)$$

The βs are the weights or regression coefficients. By using weights that minimize error in training set, the exact solution is given in a matrix form by **Equations 10** and **11** in the original and the frequency domain respectively.

$$\beta = (K - \lambda I)^{-1} P. \quad (10)$$

$$\beta_{DFT} = (K_{DFT} - \lambda_{DFT} I)^{-1} P \quad (11)$$

λ and $\lambda_{DFT}$ the ridge regression coefficients in the respective domains. I is the identity matrix and P the reference electronic properties.

IV. RESULTS

A. *1D Coulomb Signal and its Spectrogram*

In the QM7 set, the Coulomb matrix of each molecule is of size 23×23. 23 is the largest number of atoms that makes a molecule. As stated earlier, molecules with less than 23 atoms are padded with zeros so that all the Coulomb matrices have the same size 23 x 23. After algebraic manipulations, each 1D Coulomb signal and its DFT are of length 276, and are organized in matrices y and Y of size 7102×276 respectively. M=7102 is the number of molecules in the dataset, and 276 = sum(23-i), for i = 0 to 22.

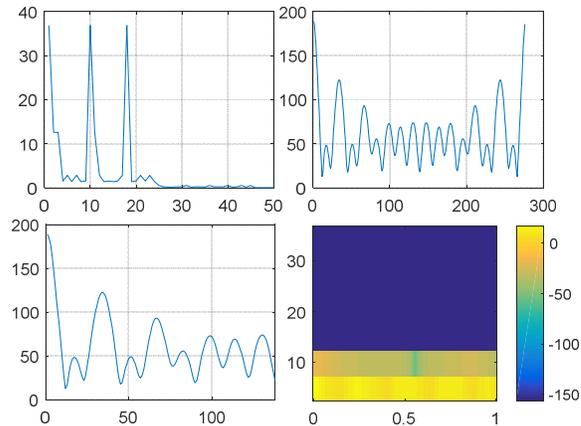

**Figure 1**. 1D Coulomb signal of propene $C_3H_6$, the absolute of its DFT transform and its spectrogram.

**Figure 1** shows the 1D Coulomb signal of propene $C_3H_6$, its DFT from k = 0 to 275, from k = 0 to 138, and its spectrogram. Spectrograms are powerful SP techniques used for visual representation of the spectrum of frequencies of a signal as it varies with time. The spectrogram corresponds to the magnitude square of the short time Fourier transform, which is obtained by applying the DFT over a sliding window of small width, thus providing a localized measure of the frequency content [10-11]. Given the ordering constraint of the 1D Coulomb signals representation of molecules, spectrograms can also be used in this case to visualize the frequency content per-se of molecules.

B. *Prediction of the Atomization Energy of Molecules*

The KRR algorithm for the prediction of electronic properties is implemented in Python. The electronic properties of molecule matrix P is of size M×1 = 7102×1, and has only one property: the atomization energy of molecules. The QM7 dataset is randomly divided into 80% training and 20% testing sets. Performance is measured using the root mean square error (RMSE), **Equation 12**, the mean absolute error (MAE), **Equation 13**, and the Pearson correlation coefficient $r_{ppe}$, **Equation 14**.

$$RMSE = \sqrt{\frac{1}{M}\sum_{m=1}^{M}(P_m - P_m^e)^2} \quad (12)$$

$$MAE = \frac{1}{M}\sum_{m=1}^{M}|P_m - P_m^e| \quad (13)$$

$$r_{PP^e} = \frac{\sum_{m=1}^{M}(P_m - \bar{P})(P_m^e - \bar{P}^e)}{\sqrt{\sum_{m=1}^{M}(P_m - \bar{P})^2}\sqrt{\sum_{m=1}^{M}(P_m^e - \bar{P}^e)^2}} \quad (14)$$

Brute force search is performed to find the optimal parameters (σ, λ) and ($\sigma_{DFT}$, $\lambda_{DFT}$) that maximizes the results in the original and frequency domain respectively. **Table I** shows the statistical performance of the results obtained, and **Table II** the optimal parameters.

TABLE I
QM7 DATASET: ATOMIZATION ENERGY PREDICTION

| Statistics | RMSE | | MAE | | $r_{PP^e}$ | |
|---|---|---|---|---|---|---|
| Domain | Original | **DFT** | Original | **DFT** | Original | DFT |
| KRR | 17.30 | **11.4** | 12.90 | **8.50** | 0.9941 | 0.9974 |
| KRR | 0.76 | **0.50** | 0.56 | **0.37** | 0.9941 | 0.9974 |

KRR: kernel ridge regression. RMSE: root mean square error. MAE: mean absolute error. RMSE and MAE are expressed in Kcal/mol in the first row and electron volts (eV) in the second row. Orig indicates the 1D signal in their original representation whereas DFT indicates the 1D signals after Discrete Fourier Transform (DFT). $r_{ppe}$ the correlation between the measured and the predicted atomization energy.

TABLE II
QM7 DATASET: OPTIMIZED PARAMETERS

| Statistics | Original | DFT |
|---|---|---|
| Kernel bandwidth (σ) | 6.905339660024878e-05 | 0.004419417382415922 |
| Regression coefficient (λ) | 1.52587890625e-05 | 0.0001220703125 |

Optimized parameters in the original and the DFT domain

**Figure 2** and **Figure 3** show the scatter plots between the measured and the predicted atomization energy in the original and frequency domain respectively.

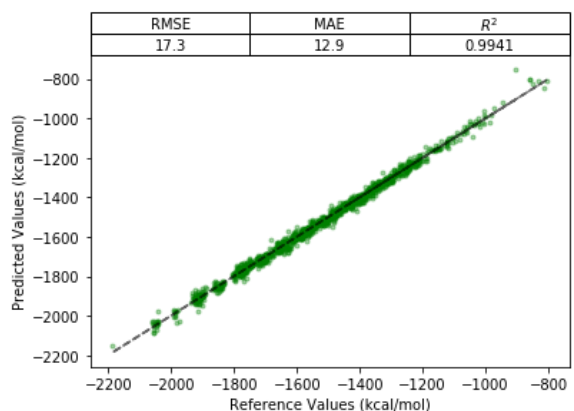

**Figure 2**. Scatter plot measured/reference versus predicted in the original domain

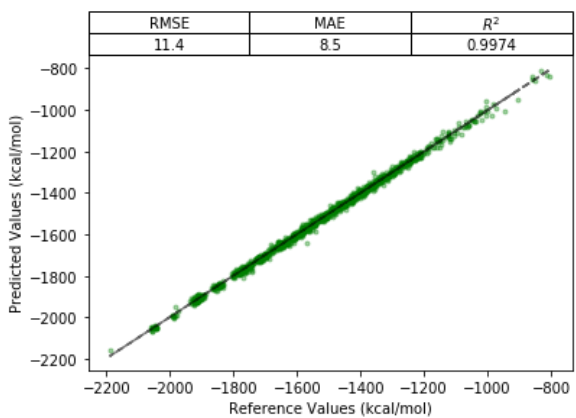

**Figure 3**. Scatter plot measured/reference versus predicted in the DFT domain

The MAE and the RMSE obtained are 4.4 Kcal/mol and 5.9 Kcal/mol lower in the frequency domain compared to the original domain and the results reported in [4,9] respectively. The MAE goes from 12.9 Kcal/mol in the original domain down to 8.5Kcal/mol in the frequency domain, whereas the RMSE goes from 17.3Kcal/mol in the original domain down to 11.4Kcal/mol in the frequency domain. Some information embedded in the original signal could have been amplified in the frequency domain. More statistical analysis should be performed to confirm this hypothesis.

## V. CONCLUSION

In this study, we presented a very preliminary view of what signal processing could do in the emerging field of quantum mechanics and machine learning data-driven models. The results obtained in this study show that the combination of reliable quantum mechanics databases with signal processing and machine learning techniques promise to be an important step towards the general goal of exploring chemical compound space for the computational design of novel and improved materials.

ACKNOWLEDGMENTS

We would like to acknowledge Shreyas Shankar, a coop-student from the University of Waterloo, who wrote part of the python code used in this study: (email: shreyas.shankar@edu.uwaterloo.ca).

REFERENCES

[1] D. Xue, P. V. Balachandran, J. Hogden, J. Theiler, D. Xue, and T. Lookman," Accelerated search for materials with targeted properties by adaptive design," Nature Communications, 2016.
[2] K.T. Butler, D.W. Davies, H. Cartwright, O. Isayev, and A. Walsh, "Machine learning for molecular and materials science," in Springer Nature, 2018, pp.547-554.
[3] L. C. Blum, J.-L. Reymond, "970 Million Druglike Small Molecules for Virtual Screening in the Chemical Universe Database GDB-13," J. Am. Chem. Soc., 131:8732, 2009.
[4] M. Rupp, A. Tkatchenko, K.-R. Müller, and O. A. von Lilienfeld, "Fast and Accurate Modeling of Molecular Atomization Energies with Machine Learning," in Phys. Rev. Lett. 2012, 108.
[5] K. Hansen, G. Montavon, F. Biegler, S. Fazli, M. Rupp, M. Scheffler, O. A. von Lilienfeld, A. Tkatchenko, and K.-R. Müller, "Assessment and Validation of Machine Learning Methods for Predicting Molecular Atomization Energies," in J. Chem. Theory Comput., 2013, 9, 3404.
[6] G. Montavon, K. Hansen, S. Fazli, M. Rupp, F. Biegler, A. Ziehe, A. Tkatchenko, O. A. von Lilienfeld, K.-R. Müller, "Learning Invariant Representations of Molecules for Atomization Energy Prediction," in Advances in Neural Information Processing Systems (NIPS), 2012.
[7] J. C. Snyder, M. Rupp, K. Hansen, K.-R. Müller, and K. Burke, "Finding Density Functionals with Machine Learning," in Phys. Rev. Lett., 2012, 108, 253002.
[8] G. Montavon, M. Rupp, V. Gobre, A. Vazquez-Mayagoitia, K. Hansen, A. Tkatchenko, K.-R. Müller, O. A. von Lilienfeld, "Machine learning of molecular electronic properties in chemicalcoumpound space," in New J. Phys. 2013, 15, 095003.
[9] M. Rupp, "Machine learning for quantum mechanics in a nutshell," in the Int. Journal of Quantum Chemistry, 2015, pp. 1058 – 1073.
[10] R.G. Lyons, "Understanding Digital Signal Processing," in (3rd ed.). Prentice Hall, 2010.
[11] B. Porat, " A Course in Digital Signal Processing," John in Wiley and Sons, 1996, pp. 27–29 and 104–105.